\RequirePackage{fix-cm}

\documentclass[english,smallextended]{svjour3}       
\smartqed  
\usepackage{graphicx}
\usepackage{amssymb}
\usepackage{amsmath}

\begin{document}

\title{Effective Hamiltonian for two interacting double dot exhange only qubits and their Controlled-NOT operations}

\author{E. Ferraro*\and M. De Michielis*\and M. Fanciulli\and E. Prati}

\institute{E. Ferraro           
              \and 
              M. De Michielis \at * Equally contributing Authors\\ Laboratorio MDM, IMM-CNR, Via Olivetti 2, I-20864 Agrate Brianza, Italy \\
              \email{elena.ferraro@mdm.imm.cnr.it} 
              \and  
              M. Fanciulli \at 
              Laboratorio MDM, IMM-CNR, Via Olivetti 2, I-20864 Agrate Brianza, Italy \at 
              Dipartimento di Scienza dei Materiali, University of Milano Bicocca, Via R. Cozzi 53, I-20126 Milano, Italy\\
              \and 
              E. Prati \at 
              Laboratorio MDM, IMM-CNR, Via Olivetti 2, I-20864 Agrate Brianza, Italy \at 
              Istituto di Fotonica e Nanotecnologia, Consiglio Nazionale delle Ricerche, Piazza Leonardo da Vinci 32, I-20133 Milano, Italy}

\date{Received: date / Accepted: date}

\maketitle

\begin{abstract}
Double dot exhange only qubit represents a promising compromise between high speed and simple fabrication in solid state implementations. A couple of interacting double dot exhange only qubits, each composed by three electrons distributed in a double quantum dot, is exploited to realize Controlled-NOT (CNOT) operations. The effective Hamiltonian model of the composite system is expressed by only exchange interactions between pairs of spins. Consequently the evolution operator has a simple form and represents the starting point for the research of sequences of operations that realize CNOT gates. Two different geometrical configurations of the pair are considered and a numerical mixed simplex and genetic algorithm is used. We compare the non physical case in which all the interactions are controllable from the external and the realistic condition in which intra-dot interactions are fixed by the geometry of the system. In the latter case, we find the CNOT sequences for both the geometrical configurations and we considered a qubit system where electrons are electrostatically confined in two quantum dots in a silicon nanowire. The effects of the geometrical sizes of the nanowire and of the gates on the fundamental parameters controlling the qubit are studied by exploiting a Spin Density Functional Theory based simulator. Consequently, CNOT gate performances are evaluated.
\PACS{03.67.Lx\and 73.21.La\and 75.10.Jm}
\end{abstract}

\section{Introduction}
\label{intro}
Several solid state physical implementations of quantum computing have been proposed both from experimental \cite{exp1,exp2,exp3,exp4,exp5,exp6} and theoretical \cite{theo1,theo2,theo3} points of view. In solid state physics many approaches have been used ranging from quantum dots, to donor-atom nuclear spins or electron spins. A characteristic common to all the scenarios is a tunable Heisenberg exchange interaction between spins to generate two-qubit quantum gate operations.

In the last decade electron spins confined in quantum dots have become an attractive basis for quantum computing because of their relatively long coherence times and potential for scaling \cite{spin1,spin2,spin3,spin4}. In the simplest proposal, single spins form the logical basis, with single qubit operations via spin resonance \cite{spin1,single}. An alternative scheme, with logical basis realized with singlet and triplet states of two spins requires inhomogeneous static magnetic field for full single-qubit control \cite{st1,st2,st3}. Using three spins to represent each qubit removes the need for an inhomogeneous field \cite{triple1}. Exchange interactions between adjacent spins suffice for all one- and two-qubit operations \cite{spin2}. Here we study a pair of double dot exchange only qubits in which three electrons are distributed in a double quantum dot \cite{shi,Shi_NatureComm2014}. The logical states are defined adopting linear combination of singlet-triplet states of a pair of electrons in one dot with single electron states of the electron in the other dot. The Schrieffer-Wolff method has already been applied to a single qubit \cite{f}. We employ the method to treat the two qubits case in order to express the Hamiltonian in terms of effective Heisenberg exchange and to exploit it to generate CNOT gates. The effective Hamiltonian explicitly express the dynamical behavior of the system and determines the conditions to realize gate operations. The most efficient sequences of interactions realizing CNOT gates have been consequently calculated in two different geometrical configurations. A numerical calculation based on a simplex method mixed with a genetic algorithm has been used.
Device designs featuring different geometrical sizes are compared by exploiting a Spin Density Function Theory (SDFT)-based simulator, presenting design rules for fast CNOT gates.

The paper is organized as follows. Section 2 is devoted to the derivation of the effective Hamiltonians for the pair of interacting qubits in the two configurations under study. Section 3 contains the main results on the sequences that realize CNOT gates in the case in which all the interactions are controllable from the external and in the realistic situation in which the intra-dot interactions are fixed by the geometry of the system. Finally in Section 4 an analysis on the gate performance in a silicon nanowire double quantum dot is presented.

\section{Effective Hamiltonian for two interacting qubits}
\begin{figure}[h]
\begin{center}
\includegraphics[width=0.5\textwidth]{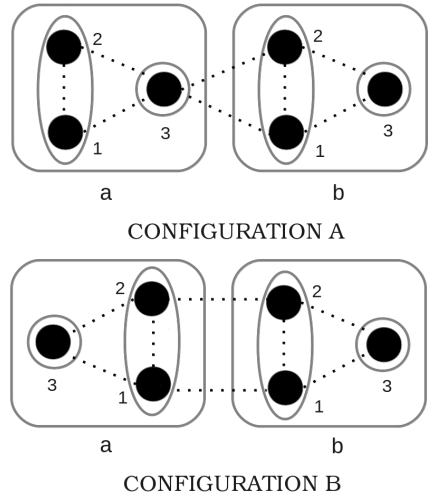}
\end{center}
\caption{Schematic of the two configurations for the couple of interacting double dot exhange only spin qubits. Dotted lines indicate the main interactions.}\label{Fig1} 
\end{figure}

In this Section we derive the Schrieffer-Wolff effective Hamiltonian to describe in a simple and intuitive way a couple of interacting qubits by combining a Hubbard-like model with a projector operator method. As a result, the Hubbard-like Hamiltonian is transformed into an equivalent expression in terms of the exchange coupling interactions between pairs of electrons. The method has been recently applied to derive the effective Hamiltonian of the single qubit in Ref.\cite{f}. The turning point in this case is represented by the way in which the two qubits are put into connection. Since the two dots composing the qubit are asymmetric, it follows that there are different possible configurations. For the purpose of our analysis we consider the two configurations shown in Fig. \ref{Fig1}. 

Generally speaking the Hamiltonian model is always expressed as the sum of three contributions:
\begin{equation}\label{H}
H=H_{a}+H_{b}+H_{ab}.
\end{equation}
The first two terms, common to all the cases under study, are the free Hamiltonians of the two qubits hereafter called $a$ and $b$, written in terms of the creation and annihilation fermionic operators $c_k^{\dagger}$ and $c_k$ respectively
\begin{equation}
H_{q}=H_{e_q}+H_{t_q}+H_{U_q}+H_{J_q}
\end{equation}
where $q=a,b$ and with 
\begin{align}
&H_{e_q}=\sum_{k=1_q,\sigma}^{3_q}\varepsilon_kc^{\dagger}_{k\sigma}c_{k\sigma}\nonumber\\
&H_{t_q}=t_{1_q3_q}\sum_{\sigma}(c^{\dagger}_{1_q\sigma}c_{3_q\sigma}+h.c.)+t_{2_q3_q}\sum_{\sigma}(c^{\dagger}_{2_q\sigma}c_{3_q\sigma}+h.c.)\nonumber\\
&H_{U_q}=\sum_{k=1_q}^{3_q}U_kn_{k\uparrow}n_{k\downarrow}+U_{1_q2_q}(n_{1_q\uparrow}+n_{1_q\downarrow})(n_{2_q\uparrow}+n_{2_q\downarrow})+\nonumber\\
&+U_{1_q3_q}(n_{1_q\uparrow}+n_{1_q\downarrow})(n_{3_q\uparrow}+n_{3_q\downarrow})+U_{2_q3_q}(n_{2_q\uparrow}+n_{2_q\downarrow})(n_{3_q\uparrow}+n_{3_q\downarrow})\nonumber\\
&H_{J_q}=H^{(1_q3_q)}_J+H^{(2_q3_q)}_J+H^{(1_q2_q)}_J
\end{align}
where $q\equiv a,\,b$ and
\begin{align}\label{hj}
&H^{(i_qj_q)}_J=-J^{(i_qj_q)}_e(n_{i_q\uparrow}n_{j_q\uparrow}+n_{i_q\downarrow}n_{j_q\downarrow})-( J^{(i_qj_q)}_ec^{\dagger}_{i_q\downarrow}c^{\dagger}_{j_q\uparrow}c_{j_q\downarrow}c_{i_q\uparrow}+\nonumber\\
&+J^{(i_qj_q)}_pc^{\dagger}_{j_q\uparrow}c^{\dagger}_{j_q\downarrow}c_{i_q\uparrow}c_{i_q\downarrow}+\sum_{k_q,\sigma}J^{(i_qj_q)}_tn_{k_q\sigma}c^{\dagger}_{i_q\bar{\sigma}}c_{j_q\bar{\sigma}}+h.c. )
\end{align}
for every pair of spins considered. $H_{e_q}$ and $H_{t_q}$ describe respectively the single electron energy level of each dot and the tunneling energy. The last two terms $H_{U_q}$ and $H_{J_q}$ constitute the intra-dot and inter-dot Coulomb interactions. More in details, the parameters appearing in Eq.(\ref{hj}) are: the spin exchange $J^{(i_qj_q)}_e$, the pair-hopping $J^{(i_qj_q)}_p$ and the occupation-modulated hopping terms $J^{(i_qj_q)}_t$.

The last term appearing in Eq.(\ref{H}) is the interaction Hamiltonian $H_{ab}$, whose explicit form depends case by case by the interaction mechanism, but always expressed as: 
\begin{equation}\label{int}
H_{ab}=H_t+H_U+H_J.
\end{equation}

In the next subsections we are going to present the final form of the effective Hamiltonians in the two cases depicted in Fig. \ref{Fig1}. 

\subsection{Configuration A}
The configuration A in Fig. \ref{Fig1} corresponds to the situation in which the right dot of the first qubit is put into direct connection with the left dot of the second qubit in such a way that the principal mechanism of interaction is between one orbital of one dot with two orbitals of the other. The interaction Hamiltonian (\ref{int}) is given by the sum of the following terms
\begin{align}
H_t=&t_{3_a1_b}\sum_{\sigma}(c^{\dagger}_{3_a\sigma}c_{1_b\sigma}+h.c.)+t_{3_a2_b}\sum_{\sigma}(c^{\dagger}_{3_a\sigma}c_{2_b\sigma}+h.c.)\nonumber\\
H_U=&U_{3_a1_b}(n_{3_a\uparrow}+n_{3_a\downarrow})(n_{1_b\uparrow}+n_{1_b\downarrow})+U_{3_a2_b}(n_{3_a\uparrow}+n_{3_a\downarrow})(n_{2_b\uparrow}+n_{2_b\downarrow})\nonumber\\
H_J=&H^{(3_a1_b)}_J+H^{(3_a2_b)}_J,
\end{align}
where $H_J$ is defined as in Eq.(\ref{hj}).

Following the same procedure reported in Ref.\cite{f}, we calculate the projected Hamiltonian. The effective Hamiltonian, concerning the low energy excitation, appears as the sum af all the exchange interactions between pairs of spins
\begin{align}\label{Heff2A}
H^{eff}&=\sum_{q=a,b}\left(J_{1_q3_q}\bold{S}_{1_q}\cdot\bold{S}_{3_q}+J_{2_q3_q}\bold{S}_{2_q}\cdot\bold{S}_{3_q}+J_{1_q2_q}\bold{S}_{1_q}\cdot\bold{S}_{2_q}\right)+\nonumber\\
&+J_{3_a1_b}\bold{S}_{3_a}\cdot\bold{S}_{1_b}+J_{3_a2_b}\bold{S}_{3_a}\cdot\bold{S}_{2_b}.
\end{align}
The effective coupling constants, considering that intra-dot Coulomb energies are larger than all the other contributions are explicitly given in Appendix A.

\subsection{Configuration B}
The second configuration B under study, (see Fig. \ref{Fig1}), corresponds to the situation in which the direct interaction is between the dots that present two orbitals. This structure create a greater number of interconnections between qubits than in configuration A. We assume that energy detuning between the double occupied quantum dots is small enough with respect to energy difference between levels $2_q$ and $1_q$ in each quantum dot. As a result, terms of the Hamiltonian containing tunnelling rates and exchange interactions between energy levels in different quantum dots with different indexes (i.e. $1_a2_b$, $2_a1_b$) are negligible. The interaction Hamiltonian terms are finally given by
\begin{align}
H_t=&t_{1_a1_b}\sum_{\sigma}(c^{\dagger}_{1_a\sigma}c_{1_b\sigma}+h.c.)+t_{2_a2_b}\sum_{\sigma}(c^{\dagger}_{2_a\sigma}c_{2_b\sigma}+h.c.)\nonumber\\
H_U=&U_{1_a1_b}(n_{1_a\uparrow}+n_{1_a\downarrow})(n_{1_b\uparrow}+n_{1_b\downarrow})+U_{1_a2_b}(n_{1_a\uparrow}+n_{1_a\downarrow})(n_{2_b\uparrow}+n_{2_b\downarrow})+\nonumber\\
&+U_{2_a1_b}(n_{2_a\uparrow}+n_{2_a\downarrow})(n_{1_b\uparrow}+n_{1_b\downarrow})+U_{2_a2_b}(n_{2_a\uparrow}+n_{2_a\downarrow})(n_{2_b\uparrow}+n_{2_b\downarrow})\nonumber\\
H_J=&H^{(1_a1_b)}_J+H^{(2_a2_b)}_J
\end{align}
where $H_J$ is defined as in Eq.(\ref{hj}). 

Analogously to the previous case, the effective Hamiltonian appears as the sum of exchange interactions:
\begin{align}\label{Heff2B}
H^{eff}&=\sum_{q=a,b}\left(J_{1_q3_q}\bold{S}_{1_q}\cdot\bold{S}_{3_q}+J_{2_q3_q}\bold{S}_{2_q}\cdot\bold{S}_{3_q}+J_{1_q2_q}\bold{S}_{1_q}\cdot\bold{S}_{2_q}\right)+\nonumber\\
&+J_{1_a1_b}\bold{S}_{1_a}\cdot\bold{S}_{1_b}+J_{1_a2_b}\bold{S}_{1_a}\cdot\bold{S}_{2_b}+J_{2_a1_b}\bold{S}_{2_a}\cdot\bold{S}_{1_b}+J_{2_a2_b}\bold{S}_{2_a}\cdot\bold{S}_{2_b}.
\end{align}
The effective coupling constants, under the assumption of larger intra-dot energies with respect to inter-dot ones are explicitly given in Appendix A.

\section{Gate sequences for exact CNOT gates}
This Section presents the gate sequences which realize CNOT operations in both configurations A and B. The effective Hamiltonians, in which only exchange interactions between pairs of spin with effective coupling constants appear, allow to represent the unitary evolution between the i-th and j-th spins with the operator
\begin{equation}\label{u}
U_{ij}(t)=e^{-\frac{i}{\hbar}tJ_{ij}\bold{S}_i\cdot\bold{S}_j}.
\end{equation}
After having introduced the mathematical background in Appendix B, we examine the simplified toy model in which all the interactions between spins are assumed controllable from the external in Appendix C. In particular we follow the procedure delineated in Ref.\cite{shi} composed by two steps: in the first one we use the central sequence found in Ref.\cite{shi} that give a gate operation that is locally equivalent to a CNOT operation; in the second step, single qubit operations, to be applied before and after the central sequence in order to obtain an exact CNOT gate, are found. 

In this Section we solve the realistic problem in which only inter-dot interactions are tunable from the external, while the intra-dot ones are fixed by the geometry of the system. To this end we adopt a search algorithm with a variable number of steps, showing how it is possible to obtain directly the exact CNOT for the configurations A and B.

\subsection{Physical fixed intra-dot interaction}
Let's now concentrate on the realistic physical situation in which the unavoidable intra-dot interactions are fixed and not tunable from the external. In fact $J_{12}$ does not depend on the tunnelling rates (see Eqs. \ref{couplingA} and \ref{couplingB}) whose values can span several order of magnitudes and strongly control $J_{13}$ and $J_{23}$. We assumed max($J_{13}$) = max($J_{23}$) = $J^{max}$ and we set a realistic constant value for $J_{12}$ = $J^{max}$/2 to model the control ineffectiveness. Moreover, the exchange interactions $J_{13}(t)$ and $J_{23}(t)$ are assumed to have instantaneous turn-on and turn-off.
In these hypothesis we derive for both configurations the sequences of unitary operations generated by exchange interactions as in Eq.(\ref{u}), whose product gives the CNOT operation in the total angular momentum basis. 

\subsubsection{Configuration A}
\begin{figure}[h]
\begin{center}
\includegraphics[width=0.6\textwidth]{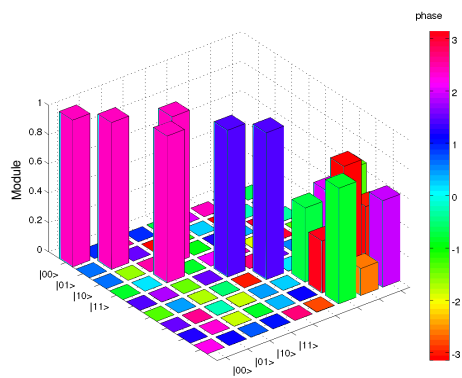}
\end{center}
\caption{Graphical representation of modulus and phase of the final transformation matrix for the CNOT gate in configuration A with gate sequence reported in Tab. \ref{Tab:SeqAndTimesCNOTv1}.}\label{v1} 
\end{figure}

The interaction sequence for an exact CNOT operation for the configuration A is calculated by using the search algorithm in the case of fixed intra-dot interaction with a final objective function value of 0.001 (see Eq. \ref{func}). The resulting sequence is reported in Tab.\ref{Tab:SeqAndTimesCNOTv1} \cite{subPRLDeMichielis}. After the CNOT sequence, the resulting transformation matrix which couples initial and final quantum states in the $9\times 9$ subspace is shown in Fig.\ref{v1}. The $4\times 4$ block in the up-left corner corresponds to the graphical representation of the CNOT matrix reported in Eq.(\ref{Eq:CNOT}). Same correspondence is obtained for the $4\times 4$ block in the $5\times 5$ subspace (not shown).
\begin{table}[th]
\caption{\label{Tab:SeqAndTimesCNOTv1}
Gate sequence implementing an exact CNOT gate for the configuration A with fixed $J_{12}$=$J^{max}$/2 in both qubits. 
Interactions in the same time step can be turned on together. The ``wait'' interaction represents only the fixed interactions $a_1 a_2$ and $b_1 b_2$ with no other interactions on.
Times are in unit of $h/J^{max}$.}
\begin{center}
\begin{tabular}{c c c|c c c}
Step & Interaction & Time & Step & Interaction & Time \\
\hline
1 & wait & 0.500 & 17 & $a_3 b_2$ & 0.299 \\ 
2 & $b_1 b_3$ & 0.866 & 18 & $a_1 a_3$ & 0.136 \\ 
3 & $a_3 b_1$ & 1.409 & 19 & $a_2 a_3$ & 0.136 \\ 
4 & $a_3 b_2$ & 0.550 & 20 & $a_3 b_1$ & 0.543 \\ 
5 & $a_1 a_3$ & 0.194 & 21 & wait & 0.907 \\ 
6 & $a_2 a_3$ & 0.194 & 22 & $b_2 b_3$ & 0.598 \\ 
7 & wait & 0.894 & 22 & $a_2 a_3$ & 0.377 \\ 
8 & $a_3 b_2$ & 0.474 & 23 & $a_1 a_3$ & 0.377 \\ 
9 & $a_1 a_3$ & 0.033 & 23 & $b_2 b_3$ & 0.736 \\ 
10 & $a_2 a_3$ & 0.279 & 24 & $a_3 b_2$ & 0.248 \\ 
11 & $a_1 a_3$ & 0.246 & 25 & wait & 0.139 \\ 
12 & $a_3 b_1$ & 0.433 & 26 & $b_2 b_3$ & 0.215 \\ 
13 & $a_3 b_2$ & 0.325 & 27 & wait & 0.254 \\ 
14 & wait & 0.210 & 28 & $a_3 b_1$ & 0.014 \\ 
15 & $a_3 b_2$ & 0.139 & 29 & $a_3 b_2$ & 0.005 \\ 
16 & $a_2 a_3$ & 1.155 & & & \\ 
\hline
\end{tabular}
\end{center}
\end{table}

\subsubsection{Configuration B}
\begin{figure}[h]
\begin{center}
\includegraphics[width=0.6\textwidth]{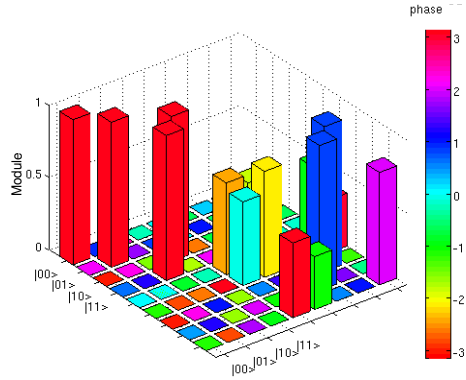}
\end{center}
\caption{Graphical representation of modulus and phase of the final transformation matrix for the CNOT gate in configuration B with gate sequence reported in Tab. \ref{Tab:SeqAndTimesCNOTv2}.}\label{v2} 
\end{figure}

For the configuration B the interaction sequence is presented in Tab.\ref{Tab:SeqAndTimesCNOTv2} \cite{subPRLDeMichielis} and the final transformation matrix, whose $4\times 4$ block in the up-left corner correctly corresponds to a CNOT matrix, is shown in Fig.\ref{v2}. Same correspondence is obtained in the $5\times 5$ subspace (not shown).
\begin{table}[th]
\caption{\label{Tab:SeqAndTimesCNOTv2}
As in Tab.\ref{Tab:SeqAndTimesCNOTv1} but for the configuration B.}
\begin{center}
\begin{tabular}{c c c|c c c}
Step & Interaction & Time & Step & Interaction & Time \\
\hline
1 & wait & 1.200 & 15 & wait & 1.604 \\ 
2 & $a_2 a_3$ & 0.238 & 16 & $a_2 b_2$ & 0.121 \\ 
3 & $a_1 a_3$ & 0.238 & 17 & wait & 1.312 \\ 
3 & $b_1 b_3$ & 0.068 & 18 & $a_2 b_2$ & 0.168 \\ 
3 & $a_2 b_2$ & 0.429 & 19 & wait & 0.400 \\ 
4 & $a_1 b_1$ & 0.439 & 20 & $b_1 b_3$ & 0.066 \\ 
4 & $b_2 b_3$ & 0.047 & 21 & $a_1 b_1$ & 1.994 \\ 
5 & $b_1 b_3$ & 0.645 & 21 & $a_2 b_2$ & 0.150 \\ 
6 & wait & 1.999 & 22 & $a_1 b_1$ & 0.268 \\ 
7 & $a_2 b_2$ & 0.242 & 23 & $b_1 b_3$ & 0.846 \\ 
8 & wait & 0.064 & 24 & wait & 0.036 \\ 
9 & $b_1 b_3$ & 0.034 & 25 & $a_1 b_1$ & 0.193 \\ 
10 & $a_1 b_1$ & 1.151 & 26 & wait & 0.294 \\ 
10 & $b_2 b_3$ & 0.006 & 27 & $b_2 b_3$ & 0.052 \\ 
11 & $a_2 b_2$ & 0.021 & 27 & $a_1 b_1$ & 0.885 \\ 
12 & $b_2 b_3$ & 0.271 & 28 & $a_2 b_2$ & 0.154 \\ 
12 & $a_1 b_1$ & 0.448 & 29 & wait & 0.092 \\ 
13 & $a_2 b_2$ & 0.111 & 30 & $b_2 b_3$ & 0.459 \\ 
14 & $b_2 b_3$ & 0.099 & 31 & wait & 0.125 \\ 
\hline
\end{tabular}
\end{center}
\end{table}

\section{Gate design and performances}
In this Section the design of the single qubit holder is presented and evaluation of the CNOT gate times are provided by using a Spin Density Function Theory (SDFT)-based simulator. 
The simulator solves the Kohn-Sham equations in the Effective Mass Approximation (EMA) with anisotropic effective masses for each couple of valleys of silicon along $\Delta$ crystallographic directions and for both spin down and spin up populations \cite{DeMichielis_APEX2012}. When the eigenstates are obtained, the spin density concentrations are calculated and the effective potentials, namely the Hartree and the LDA exchange-correlation potentials, are derived under the Local Density Approximation (LDA).
The total potential is then calculated self-consistently by solving the Poisson equation with the applied potentials from the external. The simulation ends when the error between the potential of the current iteration and that of the previous one is under a given tolerance.

\begin{figure}[t]
\begin{center}
\includegraphics[width=0.6\textwidth]{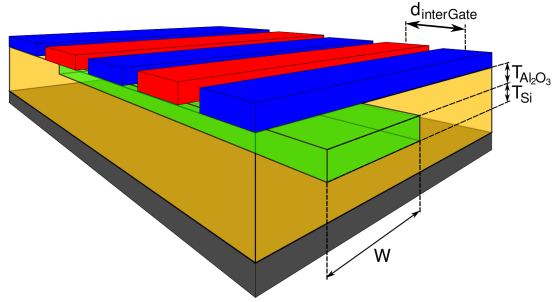}
\end{center}
\caption{The single qubit device is modeled as a silicon nanowire (in green) embedded in an insulator slab (in yellow). Accumulation gates forming the quantum dots are highlighted in red whereas the contacts controlling the inter-quantum dot electrostatic barriers are shown in blue. The back gate is shown in gray.}\label{Fig:deviceModel} 
\end{figure}

The simulated device, presented in Fig.\ref{Fig:deviceModel}, features realistic geometric sizes closer to the device (single and double quantum dots) proposed in Refs.\cite{Pierre09,Prati_Nano2012}. The device considered is a silicon nanowire featuring a rectangular section with a fixed thickness $T_{Si}$=15 nm and with a width $W$. An $Al_2O_3$ layer with thickness $T_{Al_2O_3}$=40 nm is deposed on the nanowire and 30 nm-wide $Al$ gates placed orthogonally to the nanowire direction and separated by $d_{interGate}$ are used to electrostatically confine electrons and control the inter-dot tunneling rates in the underneath silicon. 

\begin{figure}[t]
\begin{center}
\includegraphics[width=0.6\textwidth]{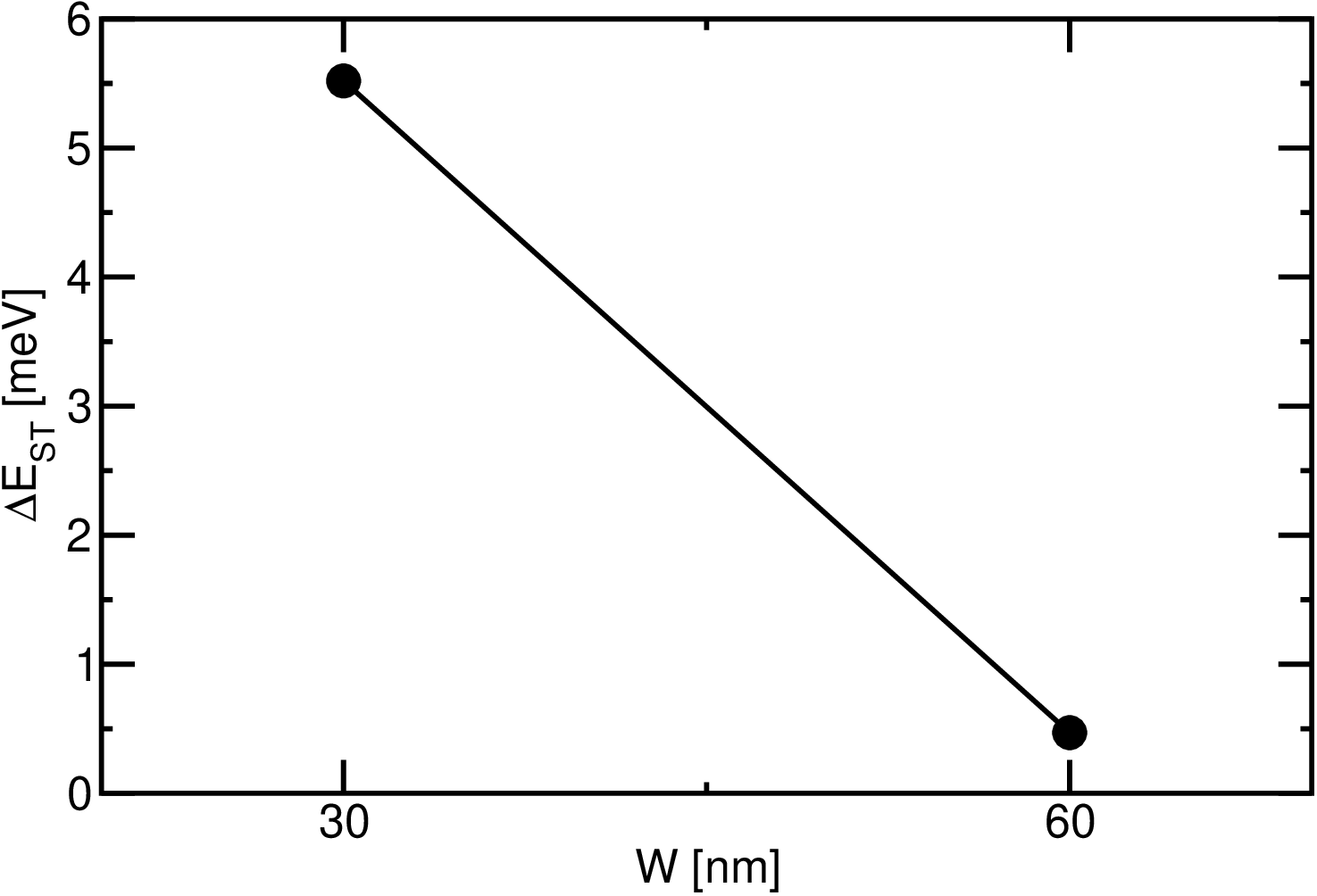}
\end{center}
\caption{Singlet-triplet energy $\Delta E_{ST}$ as a function of the nanowire width $W$.}\label{Fig:DEst-W} 
\end{figure}

The design process is the following: first of all the energy level of singlet and triplet states of the double occupied quantum dot are calculated and the singlet-triplet energy difference $\Delta E_{ST}$ is derived for two different values of $W$, 30 and 60 nm, as shown in Fig. \ref{Fig:DEst-W}. $\Delta E_{ST}$ has to be greater than the thermal energy in order to avoid unwanted transitions from singlet to triplet states. 
 
\begin{figure}[t]
\begin{center}
\includegraphics[width=0.6\textwidth]{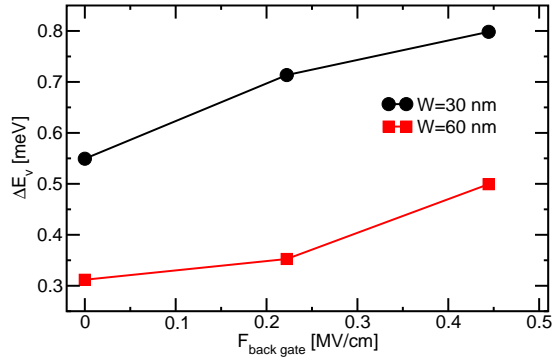}
\end{center}
\caption{Valley splitting $\Delta E_v$ as a function of the back gate electric field $F_{back}$ 
for $W$=30, 60 nm.}\label{Fig:Ev-Fback} 
\end{figure}

Then to exploit the two lowest quantum states from a single $\Delta$ valley, the valley splitting $\Delta E_{v}$ needs to be higher than the $\Delta E_{ST}$.
$\Delta E_{v}$ is defined as the difference between the two lowest energy values of fundamental couple of $\Delta$ valleys. 
$\Delta E_{v}$ is enhanced by increasing the electric field at the $Si/Al_2O_3$ interface that can be indeed obtained by polarizing negatively the back gate. 
In Fig. \ref{Fig:Ev-Fback}  the valley splitting $\Delta E_{v}$ is reported as a function of the electric field at the back interface $F_{back}$. 

\begin{figure}[t]
\begin{center}
\includegraphics[width=0.6\textwidth]{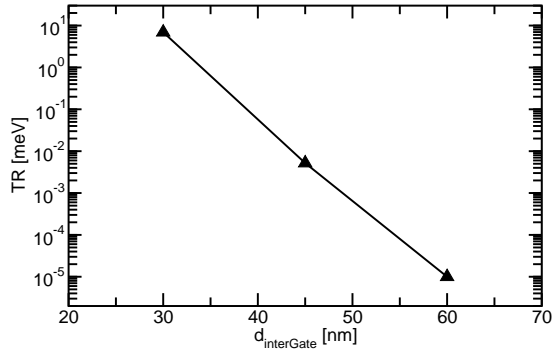}
\end{center}
\caption{Tunneling rate $TR$ as a function of the inter-gate distance $d_{interGate}$ with $W$= 60  nm.}\label{Fig:TR-dQDs} 
\end{figure}

Subsequently, the evaluation of the tunneling rate $TR$ as a function of the inter-gate distance $d_{interGate}$ is necessary to evaluate the obtainable exchange interaction values and the corresponding gate sequence time $t$. In Fig. \ref{Fig:TR-dQDs} the tunneling rate $TR$ is reported as function of $d_{interGate}$, showing that a linear reduction of $d_{interGate}$ can increase $TR$ roughly exponentially. 

Summing up the step times of the sequence (see Tables \ref{Tab:SeqAndTimesCNOTv1} and \ref{Tab:SeqAndTimesCNOTv2}) gives the total gate time in units of $h/J^{max}$. Given the tunnelling rate $TR$ and the singlet-triplet energy splitting $\Delta E_{ST}$, the maximum exchange interaction $J^{max}$ is estimated by using the formula $J^{max}$=$TR^2$/$\Delta E_{ST}$. As a result, an estimate of the total gate time can be calculated.
Simulations show that when $d_{interGate}$=45 nm, $W$=60 nm and $F_{back}$=0.45 MV/cm, the CNOT sequence durations are $t^{A}_{CNOT}$=7.4 ns and $t^{B}_{CNOT}$=10.9 ns for the configuration A and B, respectively. 
From Ref. \cite{Shi_NatureComm2014} a preliminary experimental dephasing time $T_2^*$ longer than 20 ns can be extracted, suggesting that at least a CNOT operation should be performed successfully. 

\section{Conclusions}
A pair of interacting double dot exchange only qubits is exploited to realize CNOT operations. 
We express the evolution operator in a simple form with an exchange-only effective Hamiltonian model of the composite system to obtain sequences with appropriate interactions and times. 
The search for sequences in two different configurations was performed numerically developing and using a mixed simplex and genetic algorithm. 
We compare the non physical case in which all the interactions are controllable from the external and the realistic condition in which intra-dot interactions are fixed by the geometry of the system.
Imposing realistic conditions to the interactions produces CNOT sequences with a number of steps up to 31.
  
We compare qubit designs featuring different sizes by exploiting a SDFT-based simulator. Gate performances are calculated, providing CNOT gate times of $t^{A}_{CNOT}$=7.4 ns and $t^{B}_{CNOT}$=10.9 ns for the configuration A and B, respectively.

\appendix
\section{Details of the calculation of the effective Hamiltonians for the couple of interacting qubits}
In this Appendix we are going to report all the detailed expressions for the exchange coupling constants between pair of spins in both the configurations examined. The first (last) three indices inside parenthesis, $0\le i\neq j\neq k\le 2$, assuming only integer values, denote the number of electrons in each level for qubit a (b).

\subsection{Configuration A}
The coupling constants for the configuration A are given by
\begin{align}\label{couplingA}
&J_{1_q3_q}\simeq\frac{1}{\Delta E_{1_q}}4(t_{1_q3_q}-J^{(1_q3_q)}_t)^2-2J^{(1_q3_q)}_e\nonumber\\
&J_{2_q3_q}\simeq\frac{1}{\Delta E_{2_q}}4(t_{2_q3_q}-J^{(2_q3_q)}_t)^2-2J^{(2_q3_q)}_e\nonumber\\
&J_{1_q2_q}=\left(\frac{1}{\Delta E_{3_q}}+\frac{1}{\Delta E_{4_q}}\right)4J^{(1_q2_q)2}_t-2J^{(1_q2_q)}_e\nonumber\\
&J_{3_a1_b}\simeq\frac{1}{\Delta E_5}4(t_{3_a1_b}-J^{(3_a1_b)}_t)^2-2J^{(3_a1_b)}_e\nonumber\\
&J_{3_a2_b}\simeq\frac{1}{\Delta E_6}4(t_{3_a2_b}-J^{(3_a2_b)}_t)^2-2J^{(3_a2_b)}_e.
\end{align}
where the energy differences for qubits $a$ and $b$ are
\begin{align}\label{deltaa}
&\Delta E_{1_{a(b)}}=E_{(012,111)}(E_{(111,012)})-E_{(111,111)}\nonumber\\
&\Delta E_{2_{a(b)}}=E_{(102,111)}(E_{(111,102)})-E_{(111,111)}\nonumber\\
&\Delta E_{3_{a(b)}}=E_{(201,111)}(E_{(111,201)})-E_{(111,111)}\nonumber\\
&\Delta E_{4_{a(b)}}=E_{(021,111)}(E_{(111,021)})-E_{(111,111)}\nonumber\\
&\Delta E_5=E_{(112,011)}-E_{(111,111)}\nonumber\\
&\Delta E_6=E_{(112,101)}-E_{(111,111)}\nonumber\\
\end{align}
with
\begin{align}\label{a1}
E_{(ijk,111)}=&i\varepsilon_{1_a}+j\varepsilon_{2_a}+k\varepsilon_{3_a}+ijU_{1_a2_a}+ikU_{1_a3_a}+kjU_{2_a3_a}+\delta_{i2}U_{1_a}+\delta_{j2}U_{2_a}+\delta_{k2}U_{3_a}+\nonumber\\
&+\varepsilon_{1_b}+\varepsilon_{2_b}+\varepsilon_{3_b}+U_{1_b2_b}+U_{1_b3_b}+U_{2_b3_b}+\nonumber\\
&+kU_{3_a1_b}+kU_{3_a2_b}
\end{align}
\begin{align}\label{a2}
E_{(111,ijk)}=&\varepsilon_{1_a}+\varepsilon_{2_a}+\varepsilon_{3_a}+U_{1_a2_a}+U_{1_a3_a}+U_{2_a3_a}+\nonumber\\
&+i\varepsilon_{1_b}+j\varepsilon_{2_b}+k\varepsilon_{3_b}+ijU_{1_b2_b}+ikU_{1_b3_b}+kjU_{2_b3_b}+\delta_{i2}U_{1_b}+\delta_{j2}U_{2_b}+\delta_{k2}U_{3_b}+\nonumber\\
&+iU_{3_a1_b}+iU_{3_a2_b}
\end{align}
\begin{align}\label{a3}
E_{(111,111)}=&\varepsilon_{1_a}+\varepsilon_{2_a}+\varepsilon_{3_a}+U_{1_a2_a}+U_{1_a3_a}+U_{2_a3_a}+\varepsilon_{1_b}+\varepsilon_{2_b}+\varepsilon_{3_b}+U_{1_b2_b}+U_{1_b3_b}+U_{2_b3_b}+\nonumber\\
&+U_{3_a1_b}+U_{3_a2_b}
\end{align}

\subsection{Configuration B}
The coupling constants for the configuration B are given by
\begin{align}\label{couplingB}
&J_{1_q3_q}\simeq\frac{1}{\Delta E_{1_q}}4(t_{1_q3_q}-J^{(1_q3_q)}_t)^2-2J^{(1_q3_q)}_e\nonumber\\
&J_{2_q3_q}\simeq\frac{1}{\Delta E_{2_q}}4(t_{2_q3_q}-J^{(2_q3_q)}_t)^2-2J^{(2_q3_q)}_e\nonumber\\
&J_{1_q2_q}=\left(\frac{1}{\Delta E_{3_q}}+\frac{1}{\Delta E_{4_q}}\right)4J^{(1_q2_q)2}_t-2J^{(1_q2_q)}_e\nonumber\\
&J_{1_a1_b}\simeq-2J^{(1_A1_B)}_e\nonumber\\
&J_{1_a2_b}=0\nonumber\\
&J_{2_a1_b}=0\nonumber\\
&J_{2_a2_b}\simeq-2J^{(2_A2_B)}_e,
\end{align}
where the energy differences for qubits $a$ and $b$ are 
\begin{align}\label{deltab}
&\Delta E_{1_{a(b)}}=E_{(012,111)}(E_{(111,012)})-E_{(111,111)}\nonumber\\
&\Delta E_{2_{a(b)}}=E_{(102,111)}(E_{(111,102)})-E_{(111,111)}\nonumber\\
&\Delta E_{3_{a(b)}}=E_{(201,111)}(E_{(111,201)})-E_{(111,111)}\nonumber\\
&\Delta E_{4_{a(b)}}=E_{(021,111)}(E_{(111,021)})-E_{(111,111)}
\end{align}
with
\begin{align}
\label{b1}
E_{(ijk,111)}=&i\varepsilon_{1_a}+j\varepsilon_{2_a}+k\varepsilon_{3_a}+ijU_{1_a2_a}+ikU_{1_a3_a}+kjU_{2_a3_a}+\delta_{i2}U_{1_a}+\delta_{j2}U_{2_a}+\delta_{k2}U_{3_a}+\nonumber\\
&+\varepsilon_{1_b}+\varepsilon_{2_b}+\varepsilon_{3_b}+U_{1_b2_b}+U_{1_b3_b}+U_{2_b3_b}+\nonumber\\
&+iU_{1_a1_b}+iU_{1_a2_b}+jU_{2_a1_b}+jU_{2_a2_b}
\end{align}
\begin{align}\label{b2}
E_{(111,ijk)}=&\varepsilon_{1_a}+\varepsilon_{2_a}+\varepsilon_{3_a}+U_{1_a2_a}+U_{1_a3_a}+U_{2_a3_a}+\nonumber\\
&+i\varepsilon_{1_b}+j\varepsilon_{2_b}+k\varepsilon_{3_b}+ijU_{1_b2_b}+ikU_{1_b3_b}+kjU_{2_b3_b}+\delta_{i2}U_{1_b}+\delta_{j2}U_{2_b}+\delta_{k2}U_{3_b}+\nonumber\\
&+iU_{1_a1_b}+jU_{1_a2_b}+iU_{2_a1_b}+jU_{2_a2_b}
\end{align}
\begin{align}\label{b3}
E_{(111,111)}=&\varepsilon_{1_a}+\varepsilon_{2_a}+\varepsilon_{3_a}+U_{1_a2_a}+U_{1_a3_a}+U_{2_a3_a}+\varepsilon_{1_b}+\varepsilon_{2_b}+\varepsilon_{3_b}+U_{1_b2_b}+U_{1_b3_b}+U_{2_b3_b}+\nonumber\\
&+U_{1_a1_b}+U_{1_a2_b}+U_{2_a1_b}+U_{2_a2_b}
\end{align}

\section{Mathematical background}
In this Appendix the fundamental mathematical tools used in the following to derive the sequences that realize the CNOT gates are presented.

Let's introduce the logical basis $\{|0\rangle,|1\rangle\}$ used hereafter for each qubit. It is composed by singlet and triplet states of a pair of spins, for example the pair in the left dot, in combination with the angular momentum of the third spin, localized in the right dot. This means that the logical states are finally expressed in this way
\begin{equation}\label{01}
|0\rangle\equiv|S_0\rangle|\downarrow\rangle, \qquad |1\rangle\equiv\sqrt{\frac{1}{3}}|T_0\rangle|\downarrow\rangle-\sqrt{\frac{2}{3}}|T_-\rangle|\uparrow\rangle
\end{equation}
where $|S_0\rangle$, $|T_0\rangle$ and $|T_{\pm}\rangle$ are respectively the singlet and triplet states, whose explicit form, in terms of the eigenstates of $\sigma_z$, is here reported for completeness
\begin{equation}\label{st1}
|S_0\rangle=\frac{|\uparrow\downarrow\rangle-|\downarrow\uparrow\rangle}{\sqrt{2}}, \quad |T_0\rangle=\frac{|\uparrow\downarrow\rangle+|\downarrow\uparrow\rangle}{\sqrt{2}},\quad |T_-\rangle=|\downarrow\downarrow\rangle, \quad |T_+\rangle=|\uparrow\uparrow\rangle.
\end{equation}

The basis state of the Hilbert space containing three electron spins, representing one qubit, written in the computational basis via Clebsch-Gordan coefficients is given by:
\begin{align}\label{one}
&|1\rangle=|S_0\rangle|\uparrow\rangle\nonumber\\
&|2\rangle=|S_0\rangle|\downarrow\rangle\nonumber\\
&|3\rangle=\frac{1}{\sqrt{3}}\left(\sqrt{2}|T_+\rangle|\downarrow\rangle-|T_0\rangle|\uparrow\rangle\right)\nonumber\\
&|4\rangle=\frac{1}{\sqrt{3}}\left(|T_0\rangle|\downarrow\rangle-\sqrt{2}|T_-\rangle|\uparrow\rangle\right)\nonumber\\
&|5\rangle=|T_+\rangle|\uparrow\rangle\nonumber\\
&|6\rangle=\frac{1}{\sqrt{3}}\left(|T_+\rangle|\downarrow\rangle+\sqrt{2}|T_0\rangle|\uparrow\rangle\right)\nonumber\\
&|7\rangle=\frac{1}{\sqrt{3}}\left(\sqrt{2}|T_0\rangle|\downarrow\rangle+|T_-\rangle|\uparrow\rangle\right)\nonumber\\
&|8\rangle=|T_-\rangle|\downarrow\rangle
\end{align}
with $|S_0\rangle$, $|T_0\rangle$ and $|T_{\pm}\rangle$ defined in Eqs.(\ref{st1}). On the other hand the composite system of two qubits, that is six electron spins, is hereafter described by a nine-dimensional basis in the subspace with total angular momentum operator equal to $S=1$, $S_z=-1$ obtained composing the one qubit states in Eq.(\ref{one}) with appropriate Clebsch-Gordan coefficients:
\begin{align}\label{b9}
&|b_1^{(9)}\rangle=|2\rangle|2\rangle\nonumber\\
&|b_2^{(9)}\rangle=|2\rangle|4\rangle\nonumber\\
&|b_3^{(9)}\rangle=|4\rangle|2\rangle\nonumber\\
&|b_4^{(9)}\rangle=|4\rangle|4\rangle\nonumber\\
&|b_5^{(9)}\rangle=\frac{\sqrt{3}}{2}|1\rangle|8\rangle-\frac{1}{2}|2\rangle|7\rangle\nonumber\\
&|b_6^{(9)}\rangle=\frac{\sqrt{3}}{2}|3\rangle|8\rangle-\frac{1}{2}|4\rangle|7\rangle\nonumber\\
&|b_7^{(9)}\rangle=-\frac{\sqrt{3}}{2}|8\rangle|1\rangle+\frac{1}{2}|7\rangle|2\rangle\nonumber\\
&|b_8^{(9)}\rangle=-\frac{\sqrt{3}}{2}|8\rangle|3\rangle+\frac{1}{2}|7\rangle|4\rangle\nonumber\\
&|b_9^{(9)}\rangle=\frac{1}{2}\sqrt{\frac{6}{5}}\left(|6\rangle|8\rangle+|8\rangle|6\rangle\right)-\sqrt{\frac{2}{5}}|7\rangle|7\rangle.
\end{align}
Basis vectors $|b_1^{(9)}\rangle$ - $|b_4^{(9)}\rangle$ are valid encoded states and correspond exactly to the logical basis $\{|00\rangle,|01\rangle,|10\rangle,|11\rangle\}$ in which the CNOT gate has the usual form
\begin{equation} \label{Eq:CNOT}
CNOT=\left(\begin{array}{cccc}
1 & 0 & 0 & 0 \\
0 & 1 & 0 & 0 \\
0 & 0 & 0 & 1 \\
0 & 0 & 1 & 0
\end{array}\right),
\end{equation}
basis vectors $|b_5^{(9)}\rangle$ - $|b_9^{(9)}\rangle$ are leaked states.

The objective function used in the fixed-step genetic algorithm to derive single qubit operation is:
\begin{equation}\label{func0}
f_{CNOT}=\sqrt{1-\frac{1}{4}|U^{(9)}_{(1,1)}+U^{(9)}_{(2,2)}+U^{(9)}_{(3,4)}+U^{(9)}_{(4,3)}|},
\end{equation}
where $U^{(9)}_{(i,j)}$ are the matrix elements of the CNOT in the encoded states in the $9\times 9$ subspace. The objective function is exactly equal to zero when all the $U^{(9)}$ entering into Eq.(\ref{func0}) have modulus 1 and a common phase in each subspace. 

To study the realistic situation in which intra-dot interactions are fixed by the geometry of the system, similarly to Ref.\cite{fong} a search algorithm with a variable number of time steps is developed and used. Following the procedure in Ref.\cite{fong} we adopt an objective function for the genetic algorithm that, due to the structures of exchange matrices, is confined into two subspaces for the total angular momentum operator of the composite system. The basis of the first $5\times 5$ block with total angular momentum $S=0$, $S_z=0$ is given by:
\begin{align}
&|b_1^{(5)}\rangle=\frac{1}{\sqrt{2}}\left(|1\rangle|2\rangle-|2\rangle|1\rangle\right)\nonumber\\
&|b_2^{(5)}\rangle=\frac{1}{\sqrt{2}}\left(|1\rangle|4\rangle-|2\rangle|3\rangle\right)\nonumber\\
&|b_3^{(5)}\rangle=\frac{1}{\sqrt{2}}\left(|3\rangle|2\rangle-|4\rangle|1\rangle\right)\nonumber\\
&|b_4^{(5)}\rangle=\frac{1}{\sqrt{2}}\left(|3\rangle|4\rangle-|4\rangle|3\rangle\right)\nonumber\\
&|b_5^{(5)}\rangle=\frac{1}{2}\left(|5\rangle|8\rangle-|8\rangle|5\rangle+|7\rangle|6\rangle-|6\rangle|7\rangle\right),
\end{align}
where the states on the right of the equations are given in Eq.(\ref{one}); the basis of the second block of dimension $9\times 9$ with $S=1$, $S_z=-1$ has been previously defined in Eq.(\ref{b9}). The three $9\times 9$ blocks in correspondence to $S=1$ are identical, so they need not being constrained separately. Moreover the CNOT matrix on the subspaces in correspondence to $S=2$ and $S=3$ is completely unconstrained, which is automatically satisfied with the exchange gates. 
The objective function is finally defined by 
\begin{equation}\label{func}
f_{CNOT}=\sqrt{2-\frac{1}{4}|U^{(5)}_{(1,1)}+U^{(5)}_{(2,2)}+U^{(5)}_{(3,4)}+U^{(5)}_{(4,3)}|-\frac{1}{4}|U^{(9)}_{(1,1)}+U^{(9)}_{(2,2)}+U^{(9)}_{(3,4)}+U^{(9)}_{(4,3)}|},
\end{equation}
having the same meaning of the objective function in Eq.(\ref{func0}). In the case of the CNOT operation the subspace with $S=0$ and $S=1$ have the same global phase.

\section{Toy model with all controllable interactions}
In this Appendix the CNOT sequence for the toy model with all controllable interactions is analyzed. In the first part the uncorrectness of sequence reported in Ref.\cite{shi} is demonstrated, while in the second the sequence starting from our genetic algorithm is derived.

In the following we present the sequence of exchange operations as done in Ref.\cite{shi} for configuration B, in which a locally equivalent CNOT gate is presented equipped with single qubit operations to obtain an exact CNOT.
In Fig.\ref{shi} we represent graphically the modulus and phase of the final transformation matrix presented in the supplemental material of Ref.\cite{shi} for the configuration B described by the Hamiltonian in Eq.(\ref{Heff2B}). It includes the sequence to obtain a locally equivalent CNOT and single qubit operations to obtain exactly a CNOT gate. 
\begin{figure}[h]
\begin{center}
\includegraphics[width=0.6\textwidth]{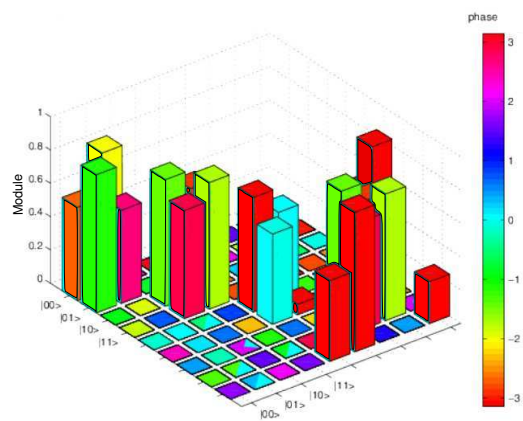}
\end{center}
\caption{Graphical representation of modulus and phase of the complete transformation erroneously called exact CNOT gate in the supplemental material of Ref.\cite{shi}.} \label{shi} 
\end{figure}

While the locally equivalent CNOT presented in the supplemental material of Ref.\cite{shi} returns $(0,1)$ Makhlin coefficients \cite{fong} as it should be, when the single qubit operations proposed are applied before and after the central sequence, the $4\times 4$ block, contrarily to they claim, does not represented an exact CNOT as Fig.\ref{shi} witnesses. 

\begin{figure}[h]
\begin{center}
\includegraphics[width=0.6\textwidth]{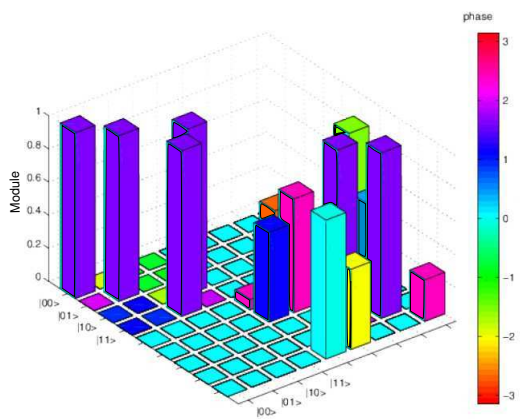}
\end{center}
\caption{Graphical representation of modulus and phase of the final transformation matrix for the exact CNOT gate in the toy model.}\label{loc} 
\end{figure}

Starting from the central gate sequence of Ref.\cite{shi} for the local equivalent CNOT in configuration B and by adopting a genetic algorithm with a fixed number of time steps to find the single qubit operations with the objective function (\ref{func0}), we obtain the exact CNOT shown in Fig. \ref{loc}.

The sequence of single qubit interactions considered is reported in Tab. \ref{Tab:SeqAndTimesSingleQubitOp}.
\begin{table}[h]
\caption{\label{Tab:SeqAndTimesSingleQubitOp}
Single qubit operations transforming the locally equivalent CNOT (LECNOT) to an exact CNOT. 
$a_1,a_2,a_3$ $(b_1,b_2,b_3)$ denote the three spins in the qubit a (b). Interaction times are in units of $h/J^{max}$. Sequences on the left and right must be applied before and after the LECNOT sequence, respectively. 
}
\begin{center}
\begin{tabular}{c c|c c}
Interactions & Time & Interactions & Time\\
 before LECNOT &  &  after LECNOT &  \\ 
\hline
$a_2 a_3$ & 0.2784 & $a_2 a_3$ & 0.3319 \\
$b_2 b_3$ & 0.9965 & $b_2 b_3$ & 0.8270 \\
$a_1 a_2$ & 0.4733 & $a_1 a_2$ & 0.5270 \\
$b_1 b_2$ & 0.2948 & $b_1 b_2$ & 0.4127 \\
$a_2 a_3$ & 0.6687 & $a_2 a_3$ & 0.7221 \\
$b_2 b_3$ & 0.5976 & $b_2 b_3$ & 0.6380 \\
\hline
\end{tabular}
\end{center}
\end{table}


\begin{thebibliography}{26}
\bibitem{sw}Schrieffer J.R., Wolff, P.A.: Relation between the Anderson and Kondo Hamiltonians. Phys. Rev. \textbf{149}, 491-492 (1966).
\bibitem{exp1}Shulman, M.D., Dial, O.E, Harvey, S.P, Bluhm, H., Umansky, V., Yacoby, A.: Demonstration of entanglement of electrostatically coupled singlet-triplet qubits. Science \textbf{336}, 202-205 (2012).
\bibitem{exp2}Johnson, A.C., Petta, J.R., Taylor, J.M., Yacoby, A., Lukin, M.D., Marcus, C.M., Hanson, M.P., Gossard, A.C.: Triplet–singlet spin relaxation via nuclei in a double quantum dot. Nature (London) \textbf{435}, 925-928 (2005).
\bibitem{exp3}Koppens, F.H.L., Folk, J.A., Elzerman, J.M., Hanson, R., Willems van Beveren, L.H., Vink, I.T., Tranitz, H.P., Wegscheider, W., Kouwenhoven, L.P., Vandersypen, L.M.K.: Control and detection of singlet-triplet mixing in a random nuclear field. Science \textbf{309}, 1346-1350 (2005).
\bibitem{exp4}Maune, B.M., Borselli, M.G., Huang, B., Ladd, T.D., Deelman, P.W., Holabird, K.S., Kiselev, A.A., Alvarado-Rodriguez, I., Ross, R.S., Schimitz, A.E., Sokolich, M., Watson, C.A., Gyure, M.F., Hunter. A.T.: Coherent singlet-triplet oscillations in a silicon-based double quantum dot. Nature (London) \textbf{481}, 344-347 (2012).
\bibitem{exp5}Bluhm, H., Foletti, S., Neder, I., Rudner, M., Mahalu, D., Umansky, V., Yacoby, A.: Dephasing time of GaAs electron-spin qubits coupled to a nuclear bath exceeding 200 $\mu$s. Nat. Phys. \textbf{7}, 109-113 (2011).
\bibitem{exp6}Tyryshkin, A.M., Tojo, S., Morton, J.J.L., Riemann, H., Abrosimov, N.V., Becker, P., Pohl, H.-J., Schenkel, T., Thewalt, M.L.W., Itoh, K.M., Lyon, S.A.: Electron spin coherence exceeding seconds in high-purity silicon. Nat. Mater. \textbf{11}, 143-147 (2012).
\bibitem{theo1}Li, R., Hu, X., You, J.Q.: Controllable exchange coupling between two singlet-triplet qubits. Phys. Rev. B \textbf{86}, 205306 (2012).
\bibitem{theo2}Coish, W.A., Loss, D.: Singlet-triplet decoherence due to nuclear spins in a double quantum dot. Phys. Rev. B \textbf{72}, 125337 (2005).
\bibitem{theo3}Shen, S.Q., Wang, Z.D.: Phase separation and charge ordering in doped manganite perovskites: Projection perturbation and mean-field approaches. Phys. Rev. B \textbf{61}, 9532-9541 (2000).
\bibitem{spin1}Loss, D., DiVincenzo, D.P.: Quantum computation with quantum dots. Phys. Rev. A \textbf{57}, 120-126 (1998).
\bibitem{spin2}DiVincenzo, D.P., Bacon, D., Kempe, J., Burkard, G., Whaley, K.: Universal quantum computation with the exchange interaction. Nature (London) \textbf{408}, 339-342 (2000).
\bibitem{spin3}Taylor, J.M., Engel, H.-A., D\"ur, W., Yacoby, A., Marcus, C.M., Zoller, P., Lukin, M.D.: Fault-tolerant architecture for quantum computation using electrically controlled semiconductor spins. Nat. Phys. \textbf{1}, 177-183 (2005).
\bibitem{spin4}Laird, 	E.A., Taylor, J.M., DiVincenzo, D.P., Marcus, C.M., Hanson, M.P., Gossard, A.C.: Coherent spin manipulation in an exchange-only qubit. Phys. Rev. B \textbf{82}, 075403 (2010). 
\bibitem{single}Vrijen, R., Yablonovitch, E., Wang, K., Jiang, H.W., Balandin, A., Roychowdhury, V., Mor, T., DiVincenzo, D.P.: Electron-spin-resonance transistors for quantum computing in silicon-germanium heterostructures. Phys. Rev. A \textbf{62}, 12306 (2000).
\bibitem{st1}Levy, J.: Universal quantum computation with spin-1/2 pairs and heisenberg exchange. Phys. Rev. Lett. \textbf{89}, 147902 (2002).
\bibitem{st2}Petta, J.R., Johnson, A.C., Taylor, J.M., Laird, E.A., Yacoby, A., Lukin, M.D., Marcus, C.M., Hanson, M.P., Gossard, A.C.: Coherent manipulation of coupled electron spins in semiconductor quantum dots. Science \textbf{309}, 2180-2184 (2005).
\bibitem{st3}Petta, J.R., Lu, H., Gossard, A.C.: A Coherent beam splitter for electronic spin states. Science \textbf{327}, 669-672 (2010).
\bibitem{triple1}Medford, J., Beil, J., Taylor, J.M., Bartlett, S.D., Doherty, A.C., Rashba, E.I., DiVincenzo, D.P., Lu, H., Gossard, A.C., Marcus, C.M.: Self-consistent measurement and state tomography of an exchange-only spin qubit. Nature Nanotechnology \textbf{8}, 654-659 (2013).
\bibitem{shi}Shi, Z., Simmons, C.B., Prance, J.R., Gamble, J.K., Koh, T.S., Shim, Y.-P., Hu, X., Savage, D.E., Lagally, M.G., Eriksson, M.A., Friesen, M., Coppersmith, S.N.: Fast hybrid silicon double-quantum-dot qubit. Phys. Rev. Lett. \textbf{108}, 140503 (2012).
\bibitem{Shi_NatureComm2014} Shi, Z., Simmons, C.B., Ward, D.R., Prance, J.R., Wu, X., Koh, T.S., Gamble, J.K., Savage, D.E., Lagally, M.G., Friesen, M., Coppersmith, S.N., Eriksson, M.A.: Fast coherent manipulation of three-electron states in a double quantum dot. Nature Communications \textbf{5}, 3020 (2014).
\bibitem{f}Ferraro, E., De Michielis, M., Mazzeo, G., Fanciulli, M., Prati, E.: Effective Hamiltonian for the hybrid double quantum dot qubit. Quantum Information Processing \textbf{13}, 1155-1173 (2014). 
\bibitem{fong}Fong, B.H., Wandzura, S.W.: Universal quantum computation and leakage reduction in the 3-qubit decoherence free subsystem. Quantum Information and Computation \textbf{11}, 1003-1018 (2011). 
\bibitem{subPRLDeMichielis} De Michielis. M., Ferraro, E., Fanciulli, M., Prati, E.: Universal set of quantum gates for double-dot exchange-only spin qubits with intradot coupling. submitted (2014).
\bibitem{DeMichielis_APEX2012} De Michielis, M., Prati, E., Fanciulli, M., Fiori, G., Iannaccone, G.: Geometrical effects on valley-orbital filling patterns in silicon quantum dots for robust qubit implementation. Applied Physics Express \textbf{5}, 124001 (2012).
\bibitem{Pierre09} Pierre, M., Wacquez, R., Roche, B., Jehl, X., Sanquer, M., Vinet, M., Prati, E., Belli, M., Fanciulli, M.: Compact silicon double and triple dots realized with only two gates. Applied Physics Letters, \textbf{95}, 242107 (2009).
\bibitem{Prati_Nano2012} Prati, E., De Michielis, M., Belli, M., Cocco, S., Fanciulli, M., Kotekar-Patil, D., Ruoff, M., Kern, D.P., Wharam, D.S., Verduijn, J., Tettamanzi, G.C., Rogge, S., Roche, B., Wacquez, R., Jehl, X., Vinet, M., Sanquer,M.: Few electron limit of n-type metal oxide semiconductor single electron transistors. Nanotechnology \textbf{23}, 215204 (2012).
\end{thebibliography}
\end{document}